# Quantum fluctuations-driven Melting Transitions in Two-dimensional Superconductors

**Authors & Affiliations:**

Dong Qiu[1,2†], Yuting Zou[3†], Chao Yang[1†], Dongxing Zheng[4], Chenhui Zhang[4], Deju Zhang[5], Yuhang Wu[1], Gaofeng Rao[1], Peng Li[1*], Yuqiao Zhou[6], Xian Jian[7], Haoran Wei[3], Zhigang Cheng[3], Xixiang Zhang[4], Yanning Zhang[5], Haiwen Liu[8*], Jingbo Qi[1,2], Yanrong Li[1], and Jie Xiong[1,2*]

[1]State Key Laboratory of Electronic Thin Films and Integrated Devices, University of Electronic Science and Technology of China, Chengdu, 610054, China

[2]School of Physics, University of Electronic Science and Technology of China, Chengdu, 610054, China

[3]Beijing National Laboratory for Condensed Matter Physics and Institute of Physics, Chinese Academy of Sciences, 100190, Beijing, China

[4]Physical Science and Engineering Division, King Abdullah University of Science and Technology, Thuwal, 23955-6900, Saudi Arabia

[5]Institute of Fundamental and Frontier Sciences, University of Electronic Science and Technology of China, Chengdu, 610054, China

[6]Key Laboratory of Green Chemistry & Technology, Ministry of Education, College of Chemistry, Sichuan University, Chengdu, 610064, China

[7]School of Materials and Energy, University of Electronic Science and Technology of China, Chengdu, 611731, China

[8]Center for Advanced Quantum Studies, Department of Physics, Beijing Normal University, Beijing, China

† These authors contributed equally to this work.

* Corresponding authors:

jiexiong@uestc.edu.cn; peng.li@uestc.edu.cn; haiwen.liu@bnu.edu.cn

**Abstract**

Quantum fluctuations are pivotal in driving quantum phase transitions, exemplified by the quantum melting of Wigner crystals into Fermi liquids in electron systems. However, their impact on superconducting systems near zero temperature, particularly in the superconductor-insulator/metal transition, remains poorly understood. In this study, through electric transport measurements on the two-dimensional (2D) superconductor $(SnS)_{1.17}NbS_2$, we demonstrate that quantum fluctuations induce vortex displacement from their mean position, leading to the quantum melting of vortex solid near zero temperature. Quantitative analysis reveals the magnetic field-induced anomalous metal originates from this quantum melting transition, with energy dissipation governed by quantum fluctuations-driven vortex displacements. Remarkably, further extending this analysis to various 2D superconductors yields the same results, and many properties of anomalous metal can be qualitatively understood within the framework of quantum melting. The connection between the quantum melting of vortex solids and dissipative anomalous metal opens a novel pathway towards understanding quantum phase transitions through vortex dynamics, providing new insights on both fields.

**Introduction**

Stemming from the uncertainty principle, quantum fluctuations play a critical role in the various physical phenomena, spanning from cosmic microwave background of the Universe [1] down to spontaneous photon emission of a single atom [2]. Thereinto, quantum phase transitions (QPT) driven by quantum fluctuations have been widely studied in cold atomic gases [3], magnets [4] and superconductors [5-7], resulting from a competition between two ground states. However, in 2D superconducting systems, this direct QPT scenario is disrupted by an intermediate bosonic anomalous metal state located in the quantum critical region of the superconductor-insulator or metal transition (SIT/SMT) [7-10]. The temperature-independent residual resistance in the anomalous metallic state has been observed in various different 2D superconductors [9,11-16], rising the natural question of whether quantum fluctuations can drive a finite dissipative state at zero temperature.

Energy dissipations in type-II superconductors generally arises from mobile vortices [17]. While the dynamics of mobile vortices at finite temperature is well understood through the thermal melting of vortex solid [18-20], their behavior at zero temperature remains poorly explored due to the unclear role of quantum fluctuations [21]. Analogous to the quantum melting of Wigner crystal in electron systems [22,23], the quantum fluctuations in superconductors may also melt vortex solids at zero temperature. This quantum melting generates mobile vortex, leading to energy dissipation at zero temperature with a subtle connection to the elusive anomalous metallic state [7]. However, experimental evidence for the quantum melting of vortex solid is scarce, and the role of quantum fluctuations in driving dissipative state remains largely uncharted.

In this study, based on the experimental measurements of the field-tuned SMT in a 2D superconductor, $(SnS)_{1.17}NbS_2$, we demonstrate that the quantum melting of vortex solid dominate the zero-temperature transport property. By identifying a linear correlation between the mobile vortex density in the anomalous metal and the quantum fluctuations-driven vortex displacements, we show that the transition towards anomalous metal is a quantum melting transition. Extending this analysis to other reported field-tuned SMTs reveals a similar relation, suggesting that field-induced anomalous metals universally originate from the quantum melting of vortex solids. The framework of quantum melting provides new insights on many puzzling properties of anomalous metal, enriching the understanding of the interplay between quantum fluctuations and exotic quantum phases in 2D superconductors.

## Results

**Two-dimensional superconductivity in (SnS)$_{1.17}$NbS$_2$**

The (SnS)$_{1.17}$NbS$_2$ single crystal is fabricated by chemical vapor transport (CVT)[24,25]. Fig. 1(a) shows a cross-section of the structure imaged by high-angle annular dark-field scanning transmission electron microscopy (HAADF-STEM) with the model structure superimposed. The transition metal dichalcogenides (TMD) layer NbS$_2$ is sandwiched between incommensurate SnS layers, which is also confirmed by the single crystal X-ray diffraction (Supplementary Fig. S2 [26]). According to the density functional theory calculations, the intercalated SnS layer induces band degeneracy which leads to the rearrangement of energy levels near the Fermi level (Supplementary Fig. S3 [26]). Thin crystalline flakes can be exfoliated from the bulk by the Scotch tape method, and electric transport measurements were conducted by measuring four-probe voltages in a Hall bar-like device configuration. Fig. 1(b) illustrates the temperature dependence of resistivity $\rho_{xx}(T)$ near the superconducting transition for a 50 nm-thick (SnS)$_{1.17}$NbS$_2$ flake. The mean-field transition temperature $T_c$ was determined as the temperature when resistance reaches 50% normal state resistance ($R_N$). The upper inset of the Fig. 1(b) shows the monotonic reduction of $T_c$ with decreasing the unit-cell number N, which follows the expected trend in niobium containing transition metal dichalcogenides.

Magnetoresistance measurements were performed to investigate the dimensionality of the superconductivity in (SnS)$_{1.17}$NbS$_2$ [27]. Fig. 1(c) demonstrates the angular-dependent upper critical field $\mu_0 H_{c2}(\theta)$ follows the 2D Tinkham model [28], $(H_{c2}(\theta)\sin\theta/H_{c2\parallel})^2 + |H_{c2}(\theta)\cos\theta/H_{c2\perp}| = 1$, instead of anisotropic 3D Ginzburg-Landau model $H_{c2}(\theta) = H_{c2\parallel}/[\sin^2\theta + (H_{c2\parallel}\cos\theta/H_{c2\perp})^2]^{1/2}$. Additionally, Fig. 1(d) also indicates that the temperature dependance of the $\mu_0 H_{c2\parallel}$ and $\mu_0 H_{c2\perp}$ follows the 2D Ginzburg-Landau (GL) behavior.

$$\mu_0 H_{c2\perp} = \frac{\Phi_0}{2\pi\xi_{GL}^2(0)}\left(1 - \frac{T}{T_c}\right) \tag{1}$$

$$\mu_0 H_{c2\parallel} = \frac{\Phi_0\sqrt{12}}{2\pi\xi_{GL}(0)d_{SC}}\sqrt{1 - \frac{T}{T_c}} \tag{2}$$

where $\Phi_0$ is the flux quantum, $\xi_{GL}(0)$ is the zero-temperature GL coherence length, and $d_{SC}$ is the temperature-independent superconducting thickness. These features clearly confirm the 2D nature of the superconductivity in (SnS)$_{1.17}$NbS$_2$.

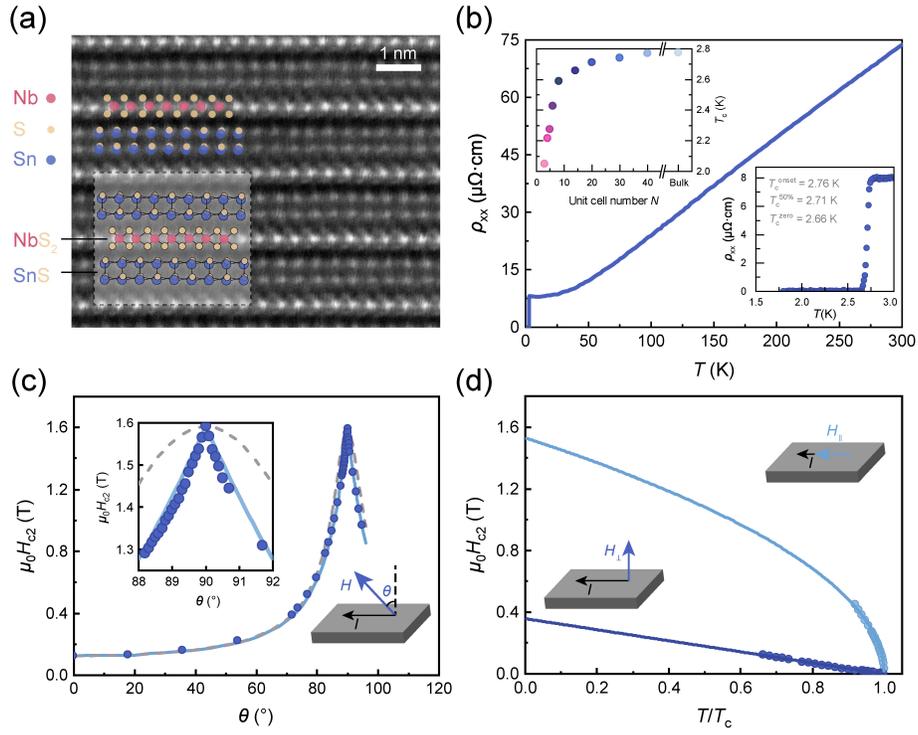

**FIG. 1. Two-dimensional Superconductivity in superlattice (SnS)$_{1.17}$NbS$_2$.** (a) HAADF-STEM image of the (SnS)$_{1.17}$NbS$_2$ taken along the [010] axis (scale bar, 1 nm). A simulation of the model structure is overlaid. Sn, S, and Nb atoms are depicted as blue, yellow, and red circles, respectively. (b) Sheet resistance as a function of temperature $\rho_{xx}(T)$ in (SnS)$_{1.17}$NbS$_2$. Upper inset: Thickness dependence of the $T_c$ in (SnS)$_{1.17}$NbS$_2$. Lower inset: $\rho_{xx}(T)$ near the superconducting transition. (c) Angular dependence of upper critical field $\mu_0 H_{c2}(\theta)$ measured at $T = 1.8$ K, which is defined by the field when resistance reaches $90\% R_N$. The solid curve and dash-dotted line are the theoretical representations of 2D-Tinkham model and 3D anisotropic Ginzburg-Landau model. (d) Temperature dependence of perpendicular upper critical field $\mu_0 H_{c2\perp}$ and parallel upper critical field $\mu_0 H_{c2\parallel}$. Solid curves are theoretical curves obtained from the 2D Ginzburg-Landau equations.

**Resistive transition towards field-induced anomalous metal**

The intrinsic ground state of (SnS)$_{1.17}$NbS$_2$ was investigated in heavily filtered dilution refrigerators (Supplementary Fig. S5-S6 [26]). As the dimensionless magnetic field, $b = B/B_{c2,T=0}$, exceeds a threshold, the superconductivity is suppressed and residual resistance appears near zero-temperature. Here we define this threshold as the onset magnetic field $b_{AM}$ of the anomalous metal, denotes the minimal dimensionless magnetic field required to induce detectable resistance at zero temperature. Fig. 2(a) and 2(b) present resistance transition of $R_{\text{sheet}}(T)$ under different perpendicular field of a thick ($N$=22UC) and a thin ($N$=3UC) sample, respectively. The existence of field-induced anomalous metal is clearly demonstrated by the resistance saturation approaching zero-temperature. Control experiments and discussions are conducted to exclude other trivial explanations for the saturated resistance [26].

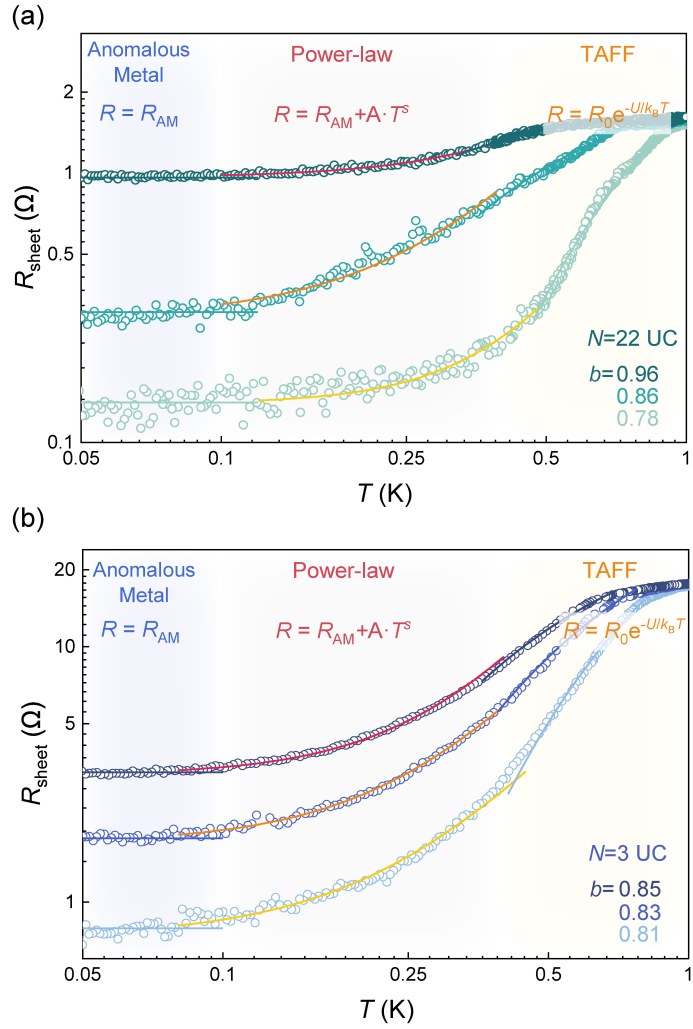

**FIG. 2. Anomalous metal, Power-law and TAFF regimes in (SnS)$_{1.17}$NbS$_2$.** (a)-(b) Three-step resistance transitions of the $R_{\text{sheet}}(T)$ of a thick (sample 1) and a thin (sample 2) (SnS)$_{1.17}$NbS$_2$ flake at various perpendicular magnetic fields. In each regime, the $R_{\text{sheet}}(T)$ is well fitted by the corresponding formulas in the figures (solid lines).

Prior to the onset of the anomalous metal, the deviation from thermally assisted flux-flow (TAFF) at low temperature has been a perplexity since its initial discovery [11,29,30]. Below we show that the exponential behavior of the TAFF is replaced by a power-law scaling at low temperature. Fig. 2(a) and 2(b) illustrate that the transition towards the anomalous metal in two independent samples both follow a three-step process. At high temperature, the resistance obeys the TAFF behavior, $R(T) = R_0 e^{-U(B)/k_B T}$, where $k_B$ is Boltzmann's constant and the $U(B)$ is the field-dependent activation energy [29,31]. Below the $T_{\text{cross}}$ (Supplementary Fig. S9 [26]), the system deviates the TAFF and enters a regime described by the power-law scaling, $R(T) = R_{\text{AM}} + A \cdot T^s$, where $R_{\text{AM}}$ is the residual resistance of the anomalous metal, $s$ is extracted exponent, and $A$ is a numeric factor. Further cooling results in the resistance saturation, $R(T) = R_{\text{AM}}$, with the onset temperature of saturation marked by $T_{\text{AM}}$ (Supplementary Fig. S10 [26]).

Fig. 3(a) shows the magnetic dependence of $T_{AM}$. Fig. 3(b) shows the extracted exponent $s$ of the relation $R(T) = R_{AM} + A \cdot T^s$ under different magnetic field, with value roughly in the range of $2.7 \pm 0.9$. This observed power-law dependance prior to the resistance saturation is reminiscent of that in the vortex glass melting transition driven by thermal fluctuations [17,18]. In the seminal work by Fisher, Fisher, and Huse [18], they derived the so-called Fisher-Fisher-Huse scaling expression of the critical behavior during the vortex glass melting, $R(T) \propto (T - T_g)^s$, where the $s$ is the critical exponent and $T_g$ is the glass melting temperature. Theoretical work and numerical simulations yield the critical exponent $s$ within the range of 2.7~8.5 [17,32], whereas the experimentally observed $s$ locates in the range of 0.6~8.0 [17,33]. The extracted exponent $s$ in our samples mostly falls in the range of 1.8~3.6, therefore, we imply the observed power-law resistivity plausibly associate with the thermal melting of vortex glass occurring at $T_g = 0$.

**Quantum melting of vortex solid at zero temperature**

The influence of quantum fluctuations gradually overwhelms thermal fluctuations as the temperature approaches zero, but how quantum fluctuations induce the anomalous metal remains an open question [5-7]. Various theoretical models have proposed that mobile vortex contribute to the finite dissipation in the anomalous metal [8,34,35], which was recently supported by Nernst measurements at low temperature [14,36]. However, the origin of the mobile vortex near zero temperature, the key to uncover the emergence of the field-induced anomalous metal, has been barely studied. Below we demonstrate that mobile vortex at zero temperature originate from the quantum melting of vortex solid [19-21]. Details of the analyzation of quantum melting is demonstrated in Section 4 of the Supplementary Materials [26].

The microscopic origin of mobile vortex near zero temperature can be quantitatively analyzed by the quantum melting of vortex solid. In the seminal theoretical work of Blatter et.al[21], they suggest that quantum fluctuations lead pinned vortices vibrate around their mean position at zero temperature, as quantified by the mean squared displacement $\langle u^2 \rangle$:

$$\frac{\langle u^2 \rangle}{\xi^2} = \frac{4}{\pi^2} \frac{R_s}{R_Q} f_0(b),$$

$$f_0(b) = \frac{1}{4} \int_0^1 dx \left\{ \ln\left(1 + \frac{\beta^2}{x^2}\right) + \ln\left(1 + \frac{\beta^2 x^2}{(x^2 + \alpha)^2}\right) \right\}. \quad (3)$$

here, $\xi$ is the superconducting coherence length, $R_s$ is the sheet resistance, $R_Q$ is the resistance quantum, the $f_0(b)$ is a function of dimensionless magnetic field $b$ with parameters $\alpha = 4 / b(1 - 0.3b)$ and $\beta = 8 / [\pi b(1-b)^2(1-0.3b)]$. Based on the Lindemann criterion [19-21], the vortex solid undergoes melting transition when the ratio between the vortex vibration displacements $\langle u^2 \rangle^{1/2}$ and vortex lattice constant $a_0$, $c = \langle u^2 \rangle^{1/2}/a_0$, exceeds the Lindemann number $c_L$. Therefore, in the scenario of quantum melting of vortex solid, the dissipative state emerges as $b > b_{AM}$, and the microscopic process of the transition can be theoretically described as follows:

$$\frac{c^2}{c_L^2} = \frac{b}{b_{AM}} \frac{f_0(b)}{f_0(b)|_{b=b_{AM}}} \quad (4)$$

The normalized vortex displacement $c^2/c_L^2$ quantifies the extent to which the vortex displacement surpasses the threshold for quantum melting, representing the degree of the transition (Supplementary Section 4 [26]). As the quantum melting transition progresses, $c^2/c_L^2$ gradually increases, and more pinned vortex becomes mobile.

To demonstrate the field-induced quantum phase transition to anomalous metal is a quantum melting transition, the above-mentioned vortex displacement should be proportional to the density of mobile vortex. Mobile vortex naturally leads dissipation [37], and the corresponding resistance $R_v = \hbar^2 \pi^2 \mu n_v / e^2$, where the $\hbar$ is the reduced Planck constant, $e$ is the electronic charge, and $\mu$ is the vortex mobility. Since (SnS)$_{1.17}$NbS$_2$ is in the dirty-limit, the vortex mobility takes the Bardeen-Stephen relation, $\mu = 2e^2 \xi_{GL}^2 R_N / \hbar^2 \pi$ [38]. Thus, the mobile vortex density can be experimentally obtained from the magnetoresistance of anomalous metal by $n_v(B) = R_{AM}(B) B_{c2,T=0} / \Phi_0 R_N$. Therefore, the proportion of mobile vortex in the anomalous metal is experimentally characterized by the relation:

$$\frac{n_v}{n_0} = \frac{R_{AM}(b)}{R_N} \frac{1}{b} \quad (5)$$

with $n_0$ denoting the vortex density ($n_0 = a_0^{-2} = B/\Phi_0$ and $a_0$ is the lattice constant of the Abrikosov vortex lattice). Fig. 3(c) shows the correlation between the $n_v/n_0$ and $c^2/c_L^2$ in (SnS)$_{1.17}$NbS$_2$ samples. The mobile vortex proportion $n_v/n_0$ increases linearly with normalized vortex displacements $c^2/c_L^2$ in three independent samples. This linear correlation indicates quantum fluctuations-induced vortex displacements cause pinned vortices to become mobile, which drives the superconductor-anomalous metal transition at zero temperature. Since this transition can be described by Lindemann melting criterion, we consider the anomalous metal originates from the quantum melting of the vortex solid.

Moreover, above-mentioned analysis can also be conducted in other different systems, such as amorphous films (Mo$_{78}$Ge$_{22}$ [14]) and crystalline materials (4Ha-TaSe$_2$ [39], PdTe$_2$ [40]). Surprisingly, as shown in Fig. 3(c), even with orders of magnitude variations in $R_N$ and $B_{c2}$, all these materials show similar linear relation between the $n_v/n_0$ and $c^2/c_L^2$. This ubiquitous linear correlation among various 2D superconductors suggest that, the field-induced anomalous metal in 2D superconductors can be quantitatively described under the framework of quantum melting of the vortex solid.

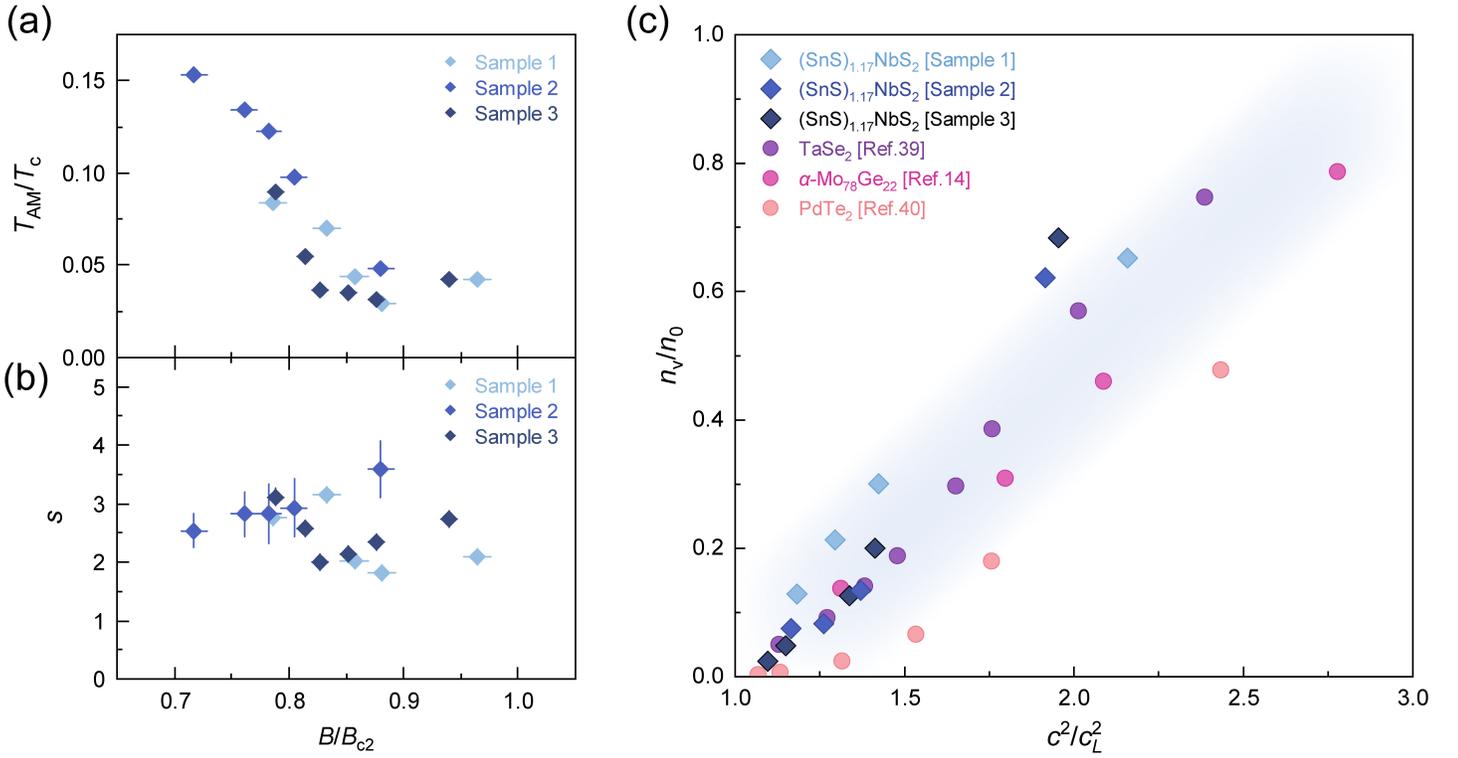

**FIG. 3. Quantum melting transitions to field-induced anomalous metals.** (a) Field-dependance of $T_{AM}$ and (b) extracted component $s$ in three independent $(SnS)_{1.17}NbS_2$ samples. (c) Correlations between the $n_v/n_0$ and $c^2/c_L^2$. The boundary of the shaded area illustrates the linear correlation across different material systems ($(SnS)_{1.17}NbS_2$ flakes, $Mo_{78}Ge_{22}$ films [14], $4Ha$-$TaSe_2$ flakes [39], and $PdTe_2$ films [40]).

**Discussion**

In Fig. 4(a), we construct a comprehensive *B-T* phase diagram to illustrate the quantum melting transition towards anomalous metal. Starting with the dissipationless vortex solid at zero temperature (blue region in Fig. 4(a)), the anomalous metal (purple region in Fig. 4(a)) can be approached via increasing magnetic field at zero temperature (blue arrow in Fig. 4(a)). Around zero temperature, the pinned vortex vibrates due to the quantum fluctuations [21]. As the vortex vibration displacement $\langle u^2 \rangle^{1/2}$ exceeds a critical value $c_L a_0$, a proportion of vortex becomes mobile and the anomalous metal emerges (Fig. 4(b)). Further increasing magnetic field leads more pinned vortex melting into mobile vortex (Fig. 4(c)). In addition, when increasing the temperature, thermal fluctuation starts depinning vortex. As the anomalous metal is 'heated up' at a fixed magnetic field (red arrow in Fig. 4(a)), the system undergoes a plausible thermal melting transition with a power-law resistivity. Further increasing temperature leads to the appearance of the conventional vortex liquid state where the TAFF behavior is recovered. Therefore, the deviation of TAFF at low temperature, known as the characteristic of the field-induced anomalous metal [11,29,30], can be understood by the occurrence of melting transitions of vortex solid.

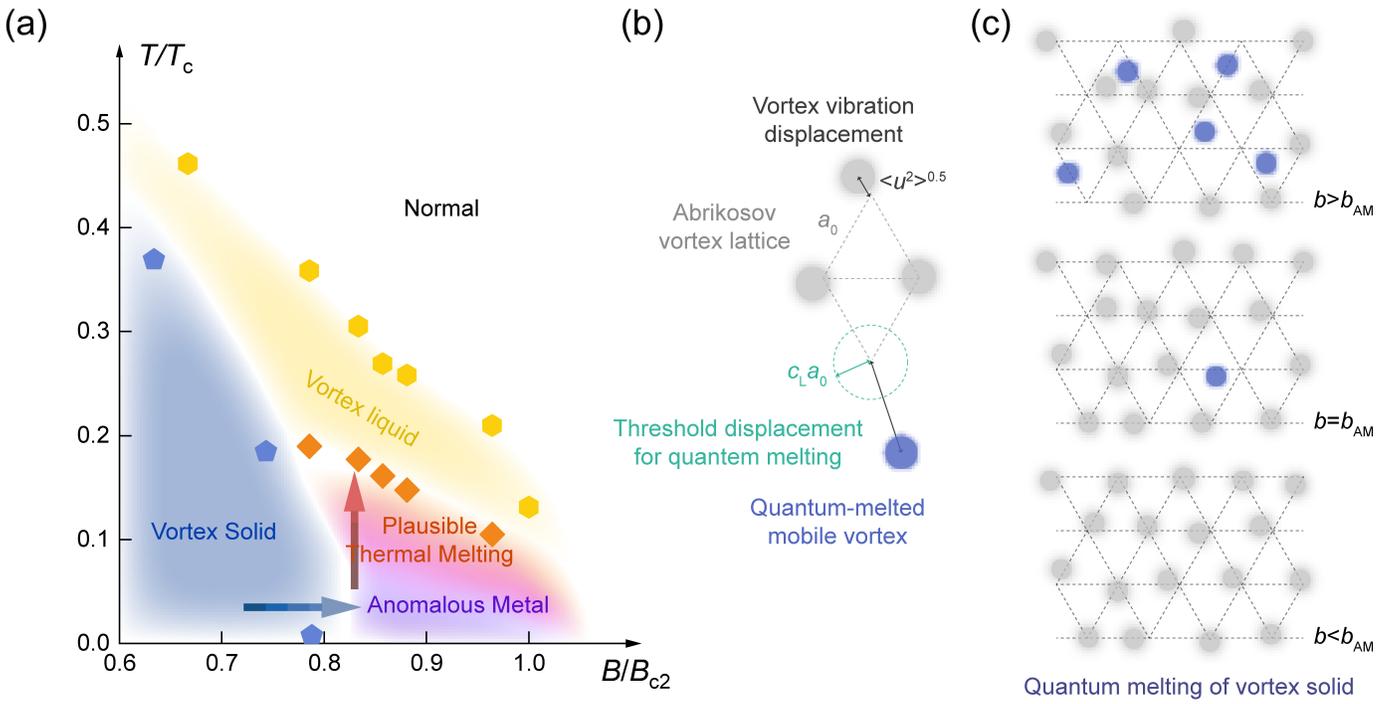

**FIG. 4. Phase diagrams and schematics of quantum melting transitions of vortex solid.** (a) The $B$-$T$ phase diagram of the $(SnS)_{1.17}NbS_2$ sample 1. Yellow hexagons divide the TAFF from the normal state. Blue pentagons denote the onset of dissipationless vortex solid state. Orange diamonds divide the TAFF with an exponential resistivity and a plausible thermal-melting regime with a power-law resistivity, marks the crossover from TAFF into the thermal melting transition. The purple region denotes the anomalous metallic state, in which the resistance is temperature-independent. Blue arrows demonstrate the quantum melting transition, respectively. Definitions of the data points and dividing lines are listed in Section 3 of the Supplementary Materials[26]. (b) The schematic of the quantum melting as described by the Lindemann criterion. Grey circles denote the pinned vortex. Blue circle denotes the quantum melted mobile vortex. Black arrows denote the vortex vibration displacement $\langle u^2 \rangle^{1/2}$. Green arrow indicates the threshold displacement $c_L a_0$ for the occurrence of quantum melting. (c) The schematic of the occurrence of quantum melting of vortex solid with increasing magnetic field, where $b_{AM}$ is the onset magnetic field of the anomalous metal.

Besides, other properties of the field-induced anomalous metal can be understood based on the above-mentioned scenario. Since the quantum fluctuation strength is irrelevant to the temperature and thermal fluctuations is negligible near zero-temperature [17,21], the mobile vortex proportion remains unchanged with decreasing temperature, and thus the resistance of anomalous metal is temperature-independent [7]. Moreover, due to rare occurrence probability of the quantum melting when magnetic field sightly exceeds $b_{AM}$, the mobile vortex density can be arbitrarily low, giving rise to a resistance of anomalous metal much lower than the $R_N$ [7]. As the magnetic field increases, more pinned vortex are dislodged by quantum fluctuations (Fig. 4(c)), leading to the giant positive magnetoresistance. Furthermore, pinned vortex are softened by quantum fluctuations-induced vortex displacements, and thus the anomalous metal is extremely sensitive to external perturbations [39,41]. For instance, softened pinned vortex can be easily dislodged by excess Lorentz force [17,42], leading to a current-dependent resistance in the anomalous metal [17,39,42].

Next, we discuss the peculiarity of the quantum melting. Fig. 4(b) illustrates this melting transition with the notion of Lindemann criterion. Compared to the thermal melting, the quantum melting is associated with the quantum fluctuations-induced time-dependent vortex motion [21]. In thermal melting, a vortex is dislodged once thermal fluctuation overcomes the pinning potential [17,18,43], hence this process is irrelevant to time. However, in the quantum melting process, the quantum fluctuations-induced vortex displacement is correlated in time domain, and a vortex can be mobile or pinned at different time [21]. This quantum fluctuations-induced motion of vortex leads to the occurrence of phase fluctuations both in spatial and temporal domain. Moreover, the inhomogeneity of the system can lead to inhomogeneous distribution of quantum fluctuation strength and the spatial variation of the quantum melting probability. Specifically, for $(SnS)_{1.17}NbS_2$, the inhomogeneity might result from the crystallographic disorder due to the misfit between the SnS and adjacent $NbS_2$ layer. However, considering the universality of the quantum melting among different systems shown in Fig. 3(c), a more general interpretation is needed to account for the origin of inhomogeneity in the QPT, such as the emergent granularity [44,45].

In conclusion, we have presented experimental evidences of the field-induced anomalous metal in a two-dimensional superconductor, $(SnS)_{1.17}NbS_2$. The transition between the superconductor and anomalous metal can be interpreted as a quantum melting of the vortex solid at zero temperature. This scenario offers a general understanding for the field-induced anomalous metal in various superconducting systems, which establishes an unprecedented link between quantum phase transition and vortex dynamics that could influence the way of thinking in both fields.

**Acknowledgments**

**Fundings:**

National Key Research and Development Program of China (Grants No. 2021YFA0718800)

National Natural Science Foundation of China (Grants No. 52021001, No. 12074056, No. 12374037 and No. U22A20132)

Sichuan Science and Technology Program (Grants No. 2021JDTD0010)

Fundamental Research Funds for the Central Universities

**Author contributions:**

J.X., H.L., and P.L. conceived the study and supervised the project. D.Q. and P.L. fabricated the samples. D.Q., Y.T. and H.W. performed the experimental measurements in dilution refrigerators. D.Q., P.L., D.Z. and C.Z. performed the angular magnetoresistance measurements. Y.W., G.R., P.L., Y.Z., X.J., and X.Z. conducted crystal growth and characterization. D.Z. and Y.Z. performed the density functional theory calculations. D.Q., H.L., P.L., Y.C. and J.X. analyzed the data. D.Q., H.L., J.X., and Y.L. wrote the manuscript with comments from J. Q. and Z.C.

**Competing interests:** The authors declare no competing interests.

**Data and materials availability:** All data are available in the main text or the supplementary materials.

**References**

[1] E. P. Tryon, Is the Universe a Vacuum Fluctuation?, Nature **246**, 396 (1973).

[2] P. W. Milonni, *An introduction to quantum optics and quantum fluctuations* (Oxford University Press, 2019).

[3] L. Bayha, M. Holten, R. Klemt, K. Subramanian, J. Bjerlin, S. M. Reimann, G. M. Bruun, P. M. Preiss, and S. Jochim, Observing the emergence of a quantum phase transition shell by shell, Nature **587**, 583 (2020).

[4] H. v. Löhneysen, A. Rosch, M. Vojta, and P. Wölfle, Fermi-liquid instabilities at magnetic quantum phase transitions, Rev. Mod. Phys. **79**, 1015 (2007).

[5] V. F. Gantmakher and V. T. Dolgopolov, Superconductor-insulator quantum phase transition, Phys.-Usp. **53**, 1 (2010).

[6] T. I. Baturina and V. M. Vinokur, Superinsulator-superconductor duality in two dimensions, Ann. Phys. (Amsterdam, Neth.) **331**, 236 (2013).

[7] A. Kapitulnik, S. A. Kivelson, and B. Spivak, Colloquium: Anomalous metals: Failed superconductors, Rev. Mod. Phys. **91**, 011002 (2019).

[8] P. Phillips and D. Dalidovich, The Elusive Bose Metal, Science **302**, 243 (2003).

[9] C. Yang, Y. Liu, Y. Wang, L. Feng, Q. He, J. Sun, Y. Tang, C. Wu, J. Xiong, W. Zhang *et al.*, Intermediate bosonic metallic state in the superconductor-insulator transition, Science **366**, 1505 (2019).

[10] C. Yang, H. Liu, Y. Liu, J. Wang, D. Qiu, S. Wang, Y. Wang, Q. He, X. Li, P. Li *et al.*, Signatures of a strange metal in a bosonic system, Nature **601**, 205 (2022).

[11] Y. Saito, Y. Kasahara, J. Ye, Y. Iwasa, and T. Nojima, Metallic ground state in an ion-gated two-dimensional superconductor, Science **350**, 409 (2015).

[12] A. W. Tsen, B. Hunt, Y. D. Kim, Z. J. Yuan, S. Jia, R. J. Cava, J. Hone, P. Kim, C. R. Dean, and A. N. Pasupathy, Nature of the quantum metal in a two-dimensional crystalline superconductor, Nat. Phys. **12**, 208 (2016).

[13] E. Sajadi, T. Palomaki, Z. Fei, W. Zhao, P. Bement, C. Olsen, S. Luescher, X. Xu, J. A. Folk, and D. H. Cobden, Gate-induced superconductivity in a monolayer topological insulator, Science **362**, 922 (2018).

[14] K. Ienaga, T. Hayashi, Y. Tamoto, S. Kaneko, and S. Okuma, Quantum Criticality inside the Anomalous Metallic State of a Disordered Superconducting Thin Film, Phys. Rev. Lett. **125**, 257001 (2020).

[15] Y. Saito, T. Nojima, and Y. Iwasa, Quantum phase transitions in highly crystalline two-dimensional superconductors, Nat. Commun. **9**, 778 (2018).

[16] X. Zhang, A. Palevski, and A. Kapitulnik, Anomalous metals: From "failed superconductor" to "failed insulator", Proc. Natl. Acad. Sci. U.S.A. **119**, e2202496119 (2022).

[17] G. Blatter, M. V. Feigel'man, V. B. Geshkenbein, A. I. Larkin, and V. M. Vinokur, Vortices in high-temperature superconductors, Rev. Mod. Phys. **66**, 1125 (1994).

[18] D. S. Fisher, M. P. A. Fisher, and D. A. Huse, Thermal fluctuations, quenched disorder, phase transitions, and transport in type-II superconductors, Phys. Rev. B **43**, 130 (1991).

[19] S. A. Khrapak, Lindemann melting criterion in two dimensions, Phys. Rev. Res. **2**, 012040 (2020).

[20] P. Lunkenheimer, A. Loidl, B. Riechers, A. Zaccone, and K. Samwer, Thermal expansion and the glass transition, Nat. Phys. **19**, 694 (2023).

[21] G. Blatter, B. Ivlev, Y. Kagan, M. Theunissen, Y. Volokitin, and P. Kes, Quantum liquid of vortices in superconductors at T=0, Phys. Rev. B **50**, 13013 (1994).

[22] M. K. Ma, K. A. Villegas Rosales, H. Deng, Y. J. Chung, L. N. Pfeiffer, K. W. West, K. W. Baldwin, R. Winkler, and M. Shayegan, Thermal and Quantum Melting Phase Diagrams for a Magnetic-Field-Induced Wigner Solid, Phys. Rev. Lett. **125**, 036601 (2020).

[23] Y. Zhou, J. Sung, E. Brutschea, I. Esterlis, Y. Wang, G. Scuri, R. J. Gelly, H. Heo, T. Taniguchi, K. Watanabe, G. Zaránd *et al.*, Bilayer Wigner crystals in a transition metal dichalcogenide heterostructure, Nature **595**, 48 (2021).

[24] G. A. Wiegers, A. Meetsma, R. J. Haange, and J. L. de Boer, Structure and physical properties of $(SnS)_{1.18}NbS_2$, "$SnNbS_3$", a compound with misfit layer structure, Mater. Res. Bull. **23**, 1551 (1988).

[25] E. E. Krasovskii, O. Tiedje, W. Schattke, J. Brandt, J. Kanzow, K. Rossnagel, L. Kipp, M. Skibowski, M. Hytha, and B. Winkler, Electronic structure and UPS of the misfit chalcogenide (SnS)NbS$_2$ and related compounds, J. Electron Spectrosc. Relat. Phenom. **114-116**, 1133 (2001).
[26] supplementary materials, **6**.
[27] M. Tinkham, *Introduction to superconductivity* (Courier Corporation, 2004).
[28] M. Tinkham, Effect of Fluxoid Quantization on Transitions of Superconducting Films, Phys. Rev. **129**, 2413 (1963).
[29] D. Ephron, A. Yazdani, A. Kapitulnik, and M. R. Beasley, Observation of Quantum Dissipation in the Vortex State of a Highly Disordered Superconducting Thin Film, Phys. Rev. Lett. **76**, 1529 (1996).
[30] Y. Qin, C. L. Vicente, and J. Yoon, Magnetically induced metallic phase in superconducting tantalum films, Phys. Rev. B **73**, 100505 (2006).
[31] M. V. Feigel'man, V. B. Geshkenbein, and A. I. Larkin, Pinning and creep in layered superconductors, Physica C **167**, 177 (1990).
[32] J. D. Reger, T. A. Tokuyasu, A. P. Young, and M. P. A. Fisher, Vortex-glass transition in three dimensions, Phys. Rev. B **44**, 7147 (1991).
[33] N. C. Yeh, D. S. Reed, W. Jiang, U. Kriplani, F. Holtzberg, A. Gupta, B. D. Hunt, R. P. Vasquez, M. C. Foote, and L. Bajuk, Scaling of vortex transport properties in high-temperature superconductors, Phys. Rev. B **45**, 5654 (1992).
[34] E. Shimshoni, A. Auerbach, and A. Kapitulnik, Transport through Quantum Melts, Phys. Rev. Lett. **80**, 3352 (1998).
[35] D. Das and S. Doniach, Existence of a Bose metal at T=0, Phys. Rev. B **60**, 1261 (1999).
[36] K. Ienaga, Y. Tamoto, M. Yoda, Y. Yoshimura, T. Ishigami, and S. Okuma, Broadened quantum critical ground state in a disordered superconducting thin film, Nat. Commun. **15**, 2388 (2024).
[37] B. I. Halperin and D. R. Nelson, Resistive transition in superconducting films, J. Low Temp. Phys. **36**, 599 (1979).
[38] Y. Kim and M. Stephen, in *Superconductivity* (Routledge, 2018), pp. 1107.
[39] Y. Xing, P. Yang, J. Ge, J. Yan, J. Luo, H. Ji, Z. Yang, Y. Li, Z. Wang, Y. Liu *et al.*, Extrinsic and Intrinsic Anomalous Metallic States in Transition Metal Dichalcogenide Ising Superconductors, Nano Lett. **21**, 7486 (2021).
[40] Y. Liu, Y. Xu, J. Sun, C. Liu, Y. Liu, C. Wang, Z. Zhang, K. Gu, Y. Tang, C. Ding *et al.*, Type-II Ising Superconductivity and Anomalous Metallic State in Macro-Size Ambient-Stable Ultrathin Crystalline Films, Nano Lett. **20**, 5728 (2020).
[41] I. Tamir, A. Benyamini, E. J. Telford, F. Gorniaczyk, A. Doron, T. Levinson, D. Wang, F. Gay, B. Sacépé, J. Hone *et al.*, Sensitivity of the superconducting state in thin films, Sci. Adv. **5**, eaau3826 (2017).
[42] A. Benyamini, E. J. Telford, D. M. Kennes, D. Wang, A. Williams, K. Watanabe, T. Taniguchi, D. Shahar, J. Hone, C. R. Dean *et al*, Fragility of the dissipationless state in clean two-dimensional superconductors, Nat. Phys. **15**, 947 (2019).
[43] A. Houghton, R. A. Pelcovits, and A. Sudbø, Flux lattice melting in high-Tc superconductors, Phys. Rev. B **40**, 6763 (1989).
[44] R. Ganguly, I. Roy, A. Banerjee, H. Singh, A. Ghosal, and P. Raychaudhuri, Magnetic field induced emergent inhomogeneity in a superconducting film with weak and homogeneous disorder, Phys. Rev. B **96**, 054509 (2017).
[45] B. Sacépé, M. Feigel'man, and T. M. Klapwijk, Quantum breakdown of superconductivity in low-dimensional materials, Nat. Phys. **16**, 734 (2020).